\documentclass[aip,amsmath,amssymb,reprint,footinbib]{revtex4-2}

\usepackage{amsmath}
\usepackage{graphicx}
\usepackage{multirow}
\usepackage{array}
\usepackage{color}
\usepackage{booktabs}
\usepackage{amsfonts}
\usepackage{hyperref}
\usepackage{textcomp}
\usepackage{ulem}
\usepackage[utf8]{inputenc}
\usepackage[T1]{fontenc}
\usepackage{mathptmx}

\begin{document}

\title{A miniature evaporator for in-operando deposition of isolated atoms in a low-temperature scanning tunneling microscope}

\author{Jeongmin Oh}
\altaffiliation{These authors contributed equally to this work.}
\affiliation{Peter Gr\"unberg Institute (PGI-3), Forschungszentrum J\"ulich, 52425 J\"ulich, Germany}
\affiliation{J\"ulich Aachen Research Alliance (JARA), Fundamentals of Future Information Technology, 52425 J\"ulich, Germany}
\affiliation{Experimentalphysik IV A, RWTH Aachen University, 52074 Aachen, Germany}
\author{Hermann Osterhage}
\altaffiliation{These authors contributed equally to this work.}
\affiliation{Peter Gr\"unberg Institute (PGI-3), Forschungszentrum J\"ulich, 52425 J\"ulich, Germany}
\affiliation{J\"ulich Aachen Research Alliance (JARA), Fundamentals of Future Information Technology, 52425 J\"ulich, Germany}
\author{Vasily Cherepanov}
\affiliation{Peter Gr\"unberg Institute (PGI-3), Forschungszentrum J\"ulich, 52425 J\"ulich, Germany}
\affiliation{J\"ulich Aachen Research Alliance (JARA), Fundamentals of Future Information Technology, 52425 J\"ulich, Germany}
\affiliation{mProbes GmbH, 52428 J\"ulich, Germany}
\author{Sven Just}
\affiliation{Peter Gr\"unberg Institute (PGI-3), Forschungszentrum J\"ulich, 52425 J\"ulich, Germany}
\affiliation{J\"ulich Aachen Research Alliance (JARA), Fundamentals of Future Information Technology, 52425 J\"ulich, Germany}
\author{Denis Krylov}
\affiliation{Peter Gr\"unberg Institute (PGI-3), Forschungszentrum J\"ulich, 52425 J\"ulich, Germany}
\affiliation{J\"ulich Aachen Research Alliance (JARA), Fundamentals of Future Information Technology, 52425 J\"ulich, Germany}
\author{F. Stefan Tautz}
\affiliation{Peter Gr\"unberg Institute (PGI-3), Forschungszentrum J\"ulich, 52425 J\"ulich, Germany}
\affiliation{J\"ulich Aachen Research Alliance (JARA), Fundamentals of Future Information Technology, 52425 J\"ulich, Germany}
\affiliation{Experimentalphysik IV A, RWTH Aachen University, 52074 Aachen, Germany}
\author{Taner Esat}
\email[Corresponding author: ]{t.esat@fz-juelich.de}
\affiliation{Peter Gr\"unberg Institute (PGI-3), Forschungszentrum J\"ulich, 52425 J\"ulich, Germany}
\affiliation{J\"ulich Aachen Research Alliance (JARA), Fundamentals of Future Information Technology, 52425 J\"ulich, Germany}
\author{Ruslan Temirov}
\email[Corresponding author: ]{r.temirov@fz-juelich.de}
\affiliation{Peter Gr\"unberg Institute (PGI-3), Forschungszentrum J\"ulich, 52425 J\"ulich, Germany}
\affiliation{University of Cologne, Faculty of Mathematics and Natural Sciences, Institute of Physics II, 50937 Cologne, Germany}

\date{\today}

\begin{abstract}
Depositing dilute atomic ensembles onto cold samples is challenging in low-temperature scanning tunneling microscopes (STM) because radiation shields and restricted internal geometries often preclude a direct deposition path, particularly in instruments designed for millikelvin operation. 
We present a compact, milliwatt-range evaporation source fabricated from a commercial miniature incandescent lamp and integrated directly into a millikelvin STM head. 
The exposed tungsten filament is coated with a micrometre-thick Fe film and positioned about 1 cm from the sample. 
We deposit isolated Fe atoms onto MgO/Ag(100) while operating the microscope near 5 K. 
Evaporation increases the STM-body temperature by only about 2 K, and the same nanoscopic surface region can be readily scanned after deposition with a lateral displacement of less than 5 nm. 
Differential-conductance spectra displaying symmetric inelastic steps near $\pm$14 mV identify the deposited atoms on MgO as Fe. 
From STM images, we estimate a local deposition flux of  $1.5\times10^{-5}~\mathrm{nm^{-2}\,s^{-1}}$, corresponding to a nominal evaporator lifetime of \textasciitilde 150 h. 
The fixed evaporator enables repeated low-flux deposition without a room-temperature line of sight, the need for movable radiation shields, or mechanical evaporator access after cooldown, while preserving access to the same atomic-scale surface region before and after deposition.
\end{abstract}

\maketitle

\section{Introduction}

Cryogenic scanning tunneling microscopy (STM) provides access to the structural, electronic, vibrational, and magnetic properties of matter at the level of individual atoms and molecules. 
Many experiments rely on preparing dilute adsorbate ensembles directly on surfaces held at cryogenic temperatures, typically below \textasciitilde 10--20 K. 
At these temperatures, thermally activated surface diffusion is strongly suppressed, preventing weakly bound atoms and molecules from aggregating and thereby stabilizing isolated adsorbates for subsequent characterization and manipulation with the STM tip. 
This approach has enabled, among other achievements, spin-excitation spectroscopy of individual magnetic atoms, the assembly of coupled atomic spin structures, and the construction of atomically defined molecular devices \cite{Heinrich2004SpinFlip,Hirjibehedin2006SpinCoupling,MartinezBlanco2015Gating,Esat2018}.

\begin{figure}
\centering
\includegraphics[width=8.5 cm]{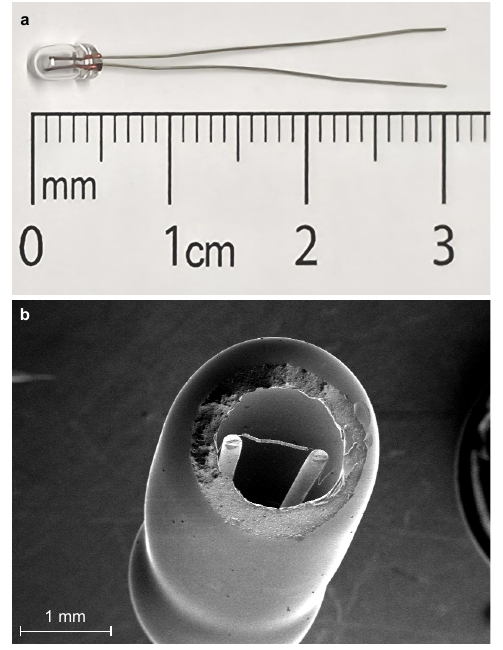}
\caption{\textbf{(a)} Photograph of the miniature incandescent lamp used to fabricate the evaporation source. \textbf{(b)} Scanning electron microscopy (SEM) image of the lamp after opening the glass bulb, exposing the tungsten filament.}
\label{fig:lamp}
\end{figure}

Introducing adsorbates into a cryogenic STM is nevertheless technically demanding. 
In a conventional arrangement, an electron-beam evaporator, effusion cell, or molecular-beam source is mounted outside the microscope and directed toward the sample through apertures in the cryostat and radiation shields \cite{Mashoff2009STM}. 
Such configurations provide a comparatively large material reservoir and may allow independent flux monitoring, but they require an unobstructed line of sight over a relatively long distance. 
Opening or penetrating the thermal shields also increases the radiative heat load on the microscope. 
These constraints become particularly severe in low-temperature STMs equipped with high-field superconducting magnets, where the available space is typically limited by the magnet bore to only a few centimetres \cite{Song2010,Assig2013,Misra2013STM,Roychowdhury2014,vonAllworden2018,Machida2018,Wong2020,Schwenk2020,Esat2021MillikelvinSTM}. 
In millikelvin instruments, the microscope is additionally surrounded by several thermal and radiation shields and is often located deep within a dilution-refrigerator\cite{Song2010,Assig2013,Misra2013STM,Roychowdhury2014,vonAllworden2018,Machida2018,Wong2020,Schwenk2020} or adiabatic-demagnetization cryostat \cite{Esat2021MillikelvinSTM}. 
As a result, these systems generally provide neither a direct deposition path from room temperature nor sufficient internal space to position a conventional evaporator close to the sample.
Evaporation concepts developed for more accessible cryogenic microscopes therefore cannot be transferred straightforwardly to this class of instruments.

Compact evaporators have previously been developed to enable deposition close to samples in cryogenic scanning probe microscopes. Lämmle \textit{et al.}\ introduced a transferable molecular evaporator based on a small resistively heated crucible \cite{Laemmle2010Evaporator}. 
The complete evaporator was transported through the UHV system and positioned near the cold sample for each deposition sequence. 
For atomic deposition, Rust \textit{et al.}\ developed a portable microevaporator from a commercial halogen lamp \cite{Rust2009Microevaporator}. 
After removal of the glass bulb, the exposed tungsten filament was loaded with the desired material and used to deposit isolated Li, Pd, and Au atoms onto samples held near 5 K. 
This evaporator likewise had to be moved into position above the sample before deposition and retracted afterwards. 
Although these approaches demonstrated controlled atomic and molecular deposition onto cryogenic samples, they required repeated mechanical access to the sample region and comparatively high heating powers.

Here, we present a compact, low-power evaporator for depositing \textit{isolated atoms} that is integrated directly into the STM head. 
Operation in the milliwatt range limits the thermal load on the microscope sufficiently to let it image the same atomic-scale surface area before and after deposition.
Because the STM remains operational and registered to the same surface region throughout the deposition cycle, we refer to this approach as \textit{in-operando} deposition.
We demonstrate the working principle by integrating two independently addressable evaporators into a millikelvin STM cooled by adiabatic demagnetization refrigeration (ADR). 
This instrument builds on the design described in Ref.~[\onlinecite{Esat2021MillikelvinSTM}].
As a second-generation system, it features a range of improvements discussed elsewhere, including a redesigned STM head that enables evaporator integration.
The fixed installation of the evaporators in very close proximity to the sample surface enables repeated deposition without direct optical access from the outside, the need for movable cryogenic shields, or an in-vacuum evaporator-transfer mechanism. 
In the following, we describe the fabrication and integration of the evaporator, characterize its performance under cryogenic STM conditions, and identify the deposited Fe atoms by inelastic electron tunneling spectroscopy.

\section{Evaporator preparation}
\label{sec:preparation}

\begin{figure}
\centering
\includegraphics[width=8.5 cm]{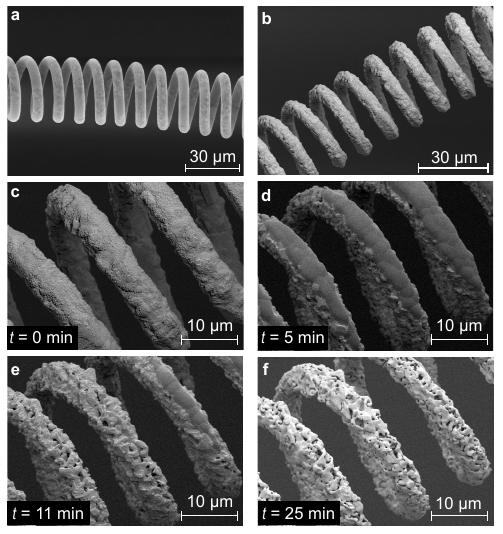}
\caption{SEM images of the evaporator filament. \textbf{(a)} Bare W filament after annealing and before Fe deposition. \textbf{(b)} Filament after coating with Fe by evaporation. \textbf{(c--f)} Higher-magnification images showing the initially fine-grained polycrystalline Fe film and its morphological evolution during resistive heating at 13 mA: \textbf{(c)} before heating, and after heating for \textbf{(d)} 5 min, \textbf{(e)} 11 min, and \textbf{(f)} 25 min.}
\label{fig:filament}
\end{figure}

We fabricated the evaporator from a commercial miniature incandescent lamp manufactured by MGG Micro-Gl\"uhlampen-Gesellschaft Menzel GmbH\footnote{Glass-bulb type: T-3/4; specified diameter range: 2.20--2.45 mm; wire-terminal standard: type 4035-00.} [Fig.~\ref{fig:lamp}(a)].
The glass bulb has a nominal diameter of 2.30 mm and a length of no more than 4.75 mm, allowing installation within the confined volume of the STM head. 
The lamp has a nominal electrical rating of 22.5 mW at 1.5 V and 15 mA, with a mean spherical candlepower of 0.002 MSCP. 
Its low-power, low-thermal-mass filament provides localized resistive heating and cools rapidly after operation, thereby limiting the thermal load on the microscope. 
The wire terminals facilitate electrical connection and mechanical mounting, while the factory-fabricated filament and supports provide a compact and reproducible evaporator geometry.

To expose the filament, as shown in Fig.~\ref{fig:lamp}(b), we removed part of the glass bulb using a rotary tool equipped with a diamond burr; fine-grit abrasive paper can be used as an alternative. 
We then cleaned the opened lamp in an aqueous detergent solution using a low-power ultrasonic bath to remove glass particles produced during grinding. 
After transfer into vacuum, we removed residual contamination from the tungsten filament by resistive annealing at 13 mA. Figure~\ref{fig:filament}(a) shows a scanning electron microscopy (SEM) image of the annealed filament.

We deposited an approximately $1~\mu\mathrm{m}$-thick Fe film onto the cleaned filament in a vacuum chamber with a base pressure of $10^{-6}$~ mbar. 
Fe with a purity of $99.99\%$ (MaTecK GmbH) was evaporated from a tungsten crucible heated by direct current. 
The new miniature evaporator charged with Fe was then transferred through air to the SEM. 
As shown in Figs.~\ref{fig:filament}(b) and \ref{fig:filament}(c), the as-deposited Fe coating exhibits a fine-grained polycrystalline morphology.

To examine the thermal evolution of the coating, we pass current through the filament inside the SEM while continuously recording its morphology. 
The image sequence in Figs.~\ref{fig:filament}(d)--\ref{fig:filament}(f) reveals pronounced coarsening of the initially fine-grained Fe film. The intermediate morphologies shown in Figs.~\ref{fig:filament}(d) and \ref{fig:filament}(e) indicate substantial thermally activated mass transport and are consistent with dewetting and coalescence, potentially accompanied by local melting. 
Next, we demonstrate the operation of the evaporator under cryogenic STM conditions by imaging the sample before and after deposition of isolated atoms.

\begin{figure}
	\centering
	\includegraphics[width=8.5 cm]{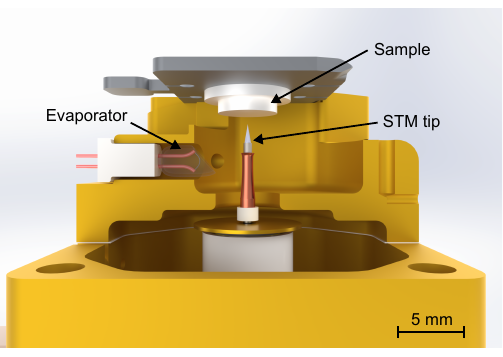}
	\caption{Three-dimensional cutaway model of the millikelvin STM head showing the integrated evaporation sources, the sample, and the STM tip.}
	\label{fig:head}
\end{figure}

\section{Evaporator operation}
\label{sec:operation}

\subsection{Installation and sample preparation}

To demonstrate the evaporator operation, we installed two identical but individually addressable Fe-loaded evaporators directly in the millikelvin STM head [Fig.~\ref{fig:head}].
The results shown here were obtained using only one evaporator.  
The evaporators were mounted before the microscope was placed under vacuum and remained fixed throughout cooldown and operation. 
The evaporators' filaments face the sample surface from approximately 1 cm. 
The evaporator arrangement fits within the restricted volume of the STM head without obstructing tip operation or sample transfer. 
The assembled STM head was then mounted on a millikelvin insert that accommodates both the microscope and the paramagnetic refrigerant used for ADR \cite{Esat2021MillikelvinSTM,esat_determining_2023}. 

After the initial cooldown, the STM head was brought into ultrahigh vacuum and cooled to 5 K without a sample installed. 
Both evaporators were then conditioned once by passing a current of 13 mA through each filament for 5 min. 
The prepared sample was inserted into the STM after the initial evaporator conditioning.

For the deposition experiments, the Ag(100) substrate was prepared in a separate chamber with a base pressure of  $2\times10^{-10}$~mbar by repeated cycles of Ar$^+$ sputtering ($P_{\mathrm{Ar}}=1.2\times10^{-7}$~mbar, $E=1$~keV, and $I_{\mathrm{ion}}=3~\mu$A) and annealing to $500\,^{\circ}\mathrm{C}$. 
MgO islands were subsequently grown by electron-beam evaporation for 10 min at an O$_2$ pressure of $1\times10^{-6}$~mbar while the sample was held at $430\,^{\circ}\mathrm{C}$ [\onlinecite{Seifert2020}].
The sample was then post-annealed for 2 min under the same oxygen pressure and cooled at $15\,^{\circ}\mathrm{C}\,\mathrm{min}^{-1}$.

After the sample reached room temperature, a submonolayer coverage of 3,4,9,10-perylenetetracarboxylic dianhydride (PTCDA) was deposited from a home-built Knudsen cell held at $305\,^{\circ}\mathrm{C}$. 
After PTCDA deposition, the sample was flashed to $100\,^{\circ}\mathrm{C}$, cooled to $-100\,^{\circ}\mathrm{C}$, and transferred into the STM.
PTCDA was not required for evaporator characterisation, but it demonstrates the applicability of the described deposition technique to experiments on surfaces coated with a variety of complex adsorbates.

\subsection{Deposition test and thermal response}

\begin{figure}
\centering
\includegraphics[width=8.5 cm]{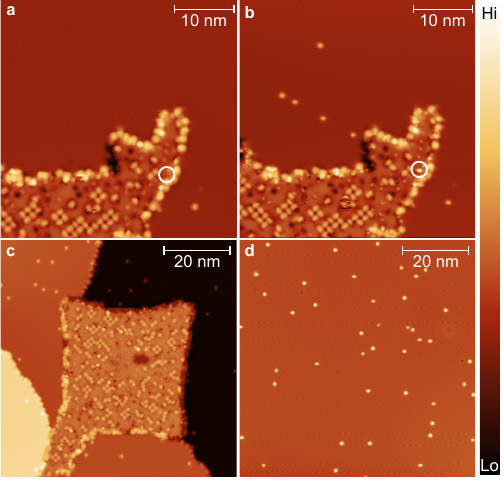}
\caption{\textbf{(a)} Constant-current STM image of the sample before Fe deposition. \textbf{(b)} The same surface area after Fe deposition. Five new atoms are visible on Ag(100), and one is visible on the MgO island (circle). \textbf{(c,d)} Representative constant-current STM images acquired after deposition, showing the Fe coverage and the cleanliness of the MgO/Ag(100) surface. Imaging parameters: $U=100$\,mV and $I=50$\,pA for (a,b), $20$\,pA for (c), and $10$\,pA for (d).}
\label{fig:stm}
\end{figure}

The constant-current STM image in Fig.~\ref{fig:stm}(a), acquired before Fe deposition, shows an irregularly shaped MgO island on a large, atomically clean Ag(100) terrace. 
Small PTCDA clusters partially cover the island and accumulate along its edges. 
The molecular ordering within these clusters is consistent with previous reports \cite{hurdax_large_2022,hurdax_integer_2025}. 
To demonstrate a controlled deposition of isolated atoms that preserves STM access to the same atomic-scale region, we deposited Fe using one of the installed evaporators and scanned the same area for comparison.

We performed the deposition at 5 K because the 15--20 mW heat load produced by the evaporator would exceed the cooling power of a millikelvin cryostat.
Additionally, we have demonstrated that ADR STM retains the same atomic-scale location upon cooling from 5 K to millikelvin temperatures\cite{Esat2021MillikelvinSTM}.
Thus, the same atomic-scale area prepared at 5 K can be studied conveniently at millikelvin temperatures. 
After recording the STM image of the test area prior to Fe deposition [Fig.~\ref{fig:stm}(a)], we disabled the feedback loop and retracted the tip by approximately 300 nm. 
We then performed Fe deposition by passing a current of $13.0\pm0.1$\,mA through the evaporator filament for a total of 9 min, split into three consecutive evaporation cycles. 
In each cycle, the filament current was ramped at $2\,\mathrm{mA}/\mathrm{min}$ to $13.0\pm0.1$\,mA and maintained at this value for 3 min.

\begin{figure}
	\centering
	\includegraphics[width=8.5 cm]{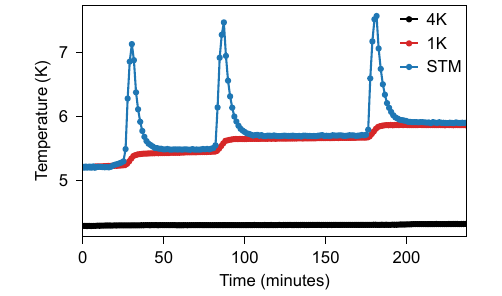}
	\caption{Temperature of the STM body (blue), 1 K pot (red) and 4 K bath (black) during three consecutive Fe evaporation cycles, each performed at $I_{\mathrm{fil}}=13$\,mA for $3$\,min.}
	\label{fig:temp}
\end{figure}

During each evaporation cycle, the STM body temperature, measured with a RuO$_x$ sensor (Entropy GmbH), rose by approximately 2 K above the 1 K pot temperature, measured with a Cernox sensor (Lakeshore), as shown in Fig.~\ref{fig:temp}. 
For the experiments reported here, we kept the 1 K pot empty and thermally decoupled it from the 4 K LHe bath.
After the filament current was switched off, the STM temperature returned within approximately 10 min to the base temperature of the 1 K pot to which it was thermally anchored \cite{Esat2021MillikelvinSTM}.
Because the 1 K pot was empty and decoupled from the 4 K bath, its temperature rose slightly after each deposition cycle.

Once the STM temperature had stabilized after the third deposition cycle, we approached the tip to the surface and acquired an image of the same area as before [Fig.~\ref{fig:stm}(b)]. 
The lateral displacement relative to the image recorded before the deposition was less than 5 nm. 
Six new point-like adsorbates appear in Figure~\ref{fig:stm}(b): five on Ag(100) and one on the MgO island.

Figures~\ref{fig:stm}(c) and \ref{fig:stm}(d) show representative regions of the surface after deposition and illustrate both the resulting Fe coverage and the overall cleanliness of the sample.
Apart from adsorbates already present before evaporation, no additional contaminant species were detected in the investigated areas. 
This indicates that exposing the evaporators to air during installation did not lead to appreciable contamination under the present conditions and that the initial conditioning procedure was sufficient to clean the evaporators for subsequent deposition. 

\subsection{Spectroscopic identification of Fe}\label{sec:iets}

\begin{figure}
\centering
\includegraphics[width=8.5 cm]{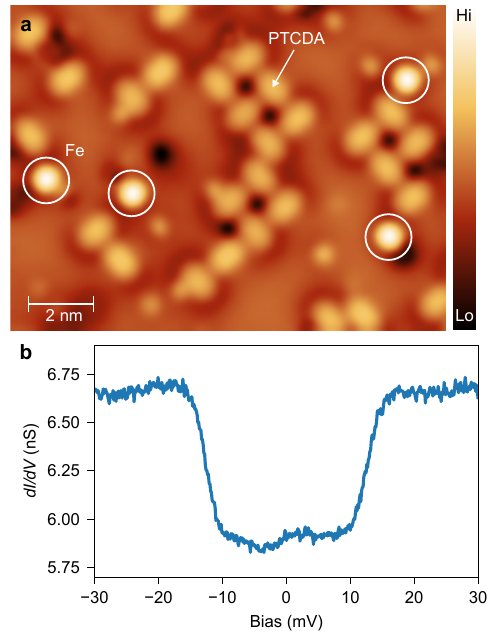}
\caption{\textbf{(a)} Constant-current STM image of the MgO/Ag(100) island shown in Fig.~\ref{fig:stm}(c), displaying four isolated Fe atoms (circled). Imaging parameters: $U=100$~mV and $I=10$~pA. \textbf{(b)} Representative differential-conductance (dI/dV) spectrum of the Fe atoms in \textbf{(a)}. Stabilization conditions: $U=30$~mV and $I=200$~pA on an Fe atom (lock-in modulation amplitude $U_{\mathrm{mod}}=1$~mV and frequency $f_{\mathrm{mod}}=589$~Hz, $T = 5.0$~K).}
\label{fig:iets}
\end{figure}

Finally, we identified the deposited atoms by inelastic electron tunneling spectroscopy (IETS).
Figure~\ref{fig:iets}(a) shows an MgO island with four isolated atoms deposited by the evaporator. 
A representative differential-conductance ($dI/dV$) spectrum of deposited atoms is shown in Fig.~\ref{fig:iets}(b).
It exhibits symmetric step-like features at approximately $U=\pm14$~mV. 
The recorded spectrum agrees well with previous measurements of individual Fe adatoms on MgO \cite{Baumann2015}.
The steps in $dI/dV$ are due to transitions from the ground-state doublet with $|m_S|=2$ to the first excited doublet with $|m_S|=1$ within the $S=2$ spin manifold driven by the inelastically tunneling electrons.
The position of each step with respect to the bias axis corresponds to the excitation energy, determined by the zero-field splitting associated with the strong out-of-plane magnetic anisotropy of Fe on MgO.

\subsection{Flux and evaporator-lifetime estimate}

To estimate the operating lifetime of the evaporation source, we compare the local flux of Fe atoms measured on the Ag(100) surface with the estimated amount of Fe initially loaded onto the filament.
The deposited-atom density is determined from clean Ag(100) regions as shown in Fig.~\ref{fig:stm}(d). A total of $N=40$ atoms is detected within an area of $A=4900~\,\mathrm{nm^2}$ after a cumulative deposition time of $t=540$\,s.  
The local flux incident on the sample is therefore

\begin{equation}
j_{\mathrm{s}} = \frac{N}{At} \approx 1.5\times10^{-5}~\mathrm{nm^{-2} s^{-1}}.
\end{equation}

Estimating the total emission rate and evaporator lifetime from this local flux requires assumptions about the evaporator geometry, angular emission distribution, and sticking probability \cite{HollandSteckelmacher1952,Ohring2002}. 
We approximate the filament as a point source emitting uniformly into the hemisphere facing the sample. 
Assuming normal incidence and a sticking coefficient of unity, the total Fe emission rate inferred from the measured sample flux is

\begin{equation}
\dot{N}_{\mathrm{em}} = 2\pi r^2 j_{\mathrm{s}} \approx 9.5\times10^{9}~\mathrm{s^{-1}},
\end{equation}

for a evaporator--sample distance of $r=1$~cm. 
If recondensation on the filament supports and surrounding components is neglected, this emission rate corresponds to the particle-loss rate of the Fe coating. 
From the SEM images, we estimate a filament-wire diameter of $d_f=5~\mu\mathrm{m}$, a coil diameter of $d_t=40~\mu\mathrm{m}$, and $n_t=60$ turns.
Assuming that an Fe layer of thickness $h=1~\mu\mathrm{m}$ covers half of the filament surface, the initial Fe volume is

\begin{equation}
V_{\mathrm{Fe}}
= \frac{1}{2}\pi^2 d_{\mathrm{f}} d_{\mathrm{t}} h n_{\mathrm{t}}
\approx 6.0\times10^4~\mu\mathrm{m^3}.
\end{equation}

Using the molar volume of Fe, $v_{\mathrm{Fe}}=7.092~\mathrm{cm^3\,mol^{-1}}=7.092\times10^{12}~\mu\mathrm{m^3\,mol^{-1}}$, the number of Fe atoms initially loaded onto the filament is

\begin{equation}
N_{\mathrm{Fe}}
= \frac{V_{\mathrm{Fe}}}{v_{\mathrm{Fe}}}N_{\mathrm{A}}
\approx 5.0\times10^{15}.
\end{equation}

The corresponding nominal operating lifetime is therefore

\begin{equation}
\tau
= \frac{N_{\mathrm{Fe}}}{\dot{N}_{\mathrm{em}}}
\approx 150~\mathrm{h}.
\end{equation}

Taking the 9-min evaporation sequence used here as a representative deposition yielding a suitable dilute Fe coverage, this operating time corresponds to approximately 1000 deposition sequences. 
This estimate assumes that the full Fe inventory is available for evaporation and that the emission rate remains constant throughout the evaporator lifetime; it should therefore be regarded as an order-of-magnitude value.

\section{Conclusion and outlook}
\label{sec:conclusion}

We have developed a compact, low-power evaporation source for depositing isolated atoms inside a cryogenic STM. The evaporator is fabricated from a commercial miniature incandescent lamp by removing part of the glass bulb and coating the exposed tungsten filament with the desired evaporant. 
Its small dimensions, factory-defined filament geometry, and milliwatt-range operation make it well suited to the restricted space and stringent thermal requirements of a cryogenic STM head. 
Because the evaporator is mounted directly on the microscope before cooldown and remains fixed during operation, deposition can be performed in operando without coarse tip retraction.

Using Fe as a model material, we demonstrated deposition onto a sample located approximately 1 cm from the filament. 
A total evaporation time of 9 min produced isolated adsorbates on both Ag(100) and MgO/Ag(100). 
During each deposition cycle, the STM-body temperature increased by only 2 K above the base temperature of the stage to which the STM was anchored and returned to the base value within about 10 min after finishing the deposition.
The same atomic-scale region could be readily scanned after deposition, with a lateral displacement of less than 5 nm. 
No additional contaminant species beyond those already present before evaporation were detected in the investigated areas, indicating that the initial evaporator-conditioning procedure was sufficient despite air exposure during installation. 
Inelastic spin-excitation spectra with conductance steps near $\pm14$~mV identified the deposited atoms on MgO as Fe.

Under a simple isotropic point-source model with hemispherical emission and unit sticking probability, the measured coverage corresponds to an Fe emission rate of approximately $9.5\times10^{9}$~per second. 
Together with the estimated Fe inventory of the filament, this yields a nominal evaporator lifetime of approximately 150 h. 
Taking the 9-min evaporation sequence used here as a representative deposition, this corresponds to roughly 1000 deposition sequences. 

The utility of the deposition technique proposed here can be extended further.
A natural next step is developing evaporators that can be exchanged in situ without warming the STM to room temperature or breaking UHV conditions.
The same compact-evaporator concept may also be adaptable to molecular deposition. 
Establishing this capability will require material-specific operating protocols that provide a stable flux while avoiding thermal decomposition. 
With these refinements, arrays of independently addressable evaporators could enable sequential deposition of several atomic and molecular species onto a sample that remains cryogenic throughout, extending atom-by-atom and molecule-by-molecule preparation to low-temperature STM architectures that currently lack practical deposition access.

\begin{acknowledgments}

V.C., S.J., D.K., F.S.T., T.E., and R.T. were supported by the German Federal Ministry of Education and Research through the funding programme "Quantum Technologies: From Basic Research to Market" under Q-NL (grant no. 13N16032). 
J.O., H.O., and T.E. were supported by ERC grant (QuSINT, 101160588, DOI: 10.3030/101160588).
Funded by the European Union. 
Views and opinions expressed are, however, those of the authors only and do not necessarily reflect those of the European Union or the European Research Council Executive Agency. 
Neither the European Union nor the granting authority can be held responsible for them.

\end{acknowledgments}

\section*{Competing interests}

V.C. and R.T. are named inventors on German patent application 2022102615185700DE, filed on 26 October 2022 by Forschungszentrum J\"ulich GmbH, entitled "Method for depositing individual atoms and/or molecules on a sample surface and apparatus for carrying out the method." This patent application relates to aspects of the technology described in this manuscript. The remaining authors declare no competing interests.

\section*{Data availability}

All data presented in this study are available on the J\"ulich Data Repository at XYZ. 

\section*{References}

\bibliography{references}

@article{esat_determining_2023,
	title = {Determining the temperature of a millikelvin scanning tunnelling microscope junction},
	volume = {6},
	pages = {81},
	number = {1},
	journal = {Communications Physics},
	author = {Esat, Taner and Yang, Xiaosheng and Mustafayev, Farhad and Soltner, Helmut and Tautz, F. Stefan and Temirov, Ruslan},
	year = {2023},
}

@article{Esat2018,
	title = {A standing molecule as a single-electron field emitter},
	volume = {558},
	pages = {573--576},
	journal = {Nature},
	author = {Esat, Taner and Friedrich, Niklas and Tautz, F. Stefan and Temirov, Ruslan},
	year = {2018},
}

@article{Rust2009Microevaporator,
  author  = {Rust, Hans-Peter and K\"onig, T. and Simon, G. H. and Nowicki, M. and Simic-Milosevic, V. and Thielsch, G. and Heyde, M. and Freund, H.-J.},
  title   = {A portable microevaporator for low temperature single atom studies by scanning tunneling and dynamic force microscopy},
  journal = {Review of Scientific Instruments},
  volume  = {80},
  number  = {11},
  pages   = {113705},
  year    = {2009},
  doi     = {10.1063/1.3266971}
}

@article{Laemmle2010Evaporator,
  author  = {L\"ammle, K. and Schwarz, A. and Wiesendanger, R.},
  title   = {Miniaturized transportable evaporator for molecule deposition inside cryogenic scanning probe microscopes},
  journal = {Review of Scientific Instruments},
  volume  = {81},
  number  = {5},
  pages   = {053902},
  year    = {2010},
  doi     = {10.1063/1.3428621}
}

@article{Mashoff2009STM,
  author  = {Mashoff, Torge and Pratzer, Marco and Morgenstern, Markus},
  title   = {A low-temperature high resolution scanning tunneling microscope with a three-dimensional magnetic vector field operating in ultrahigh vacuum},
  journal = {Review of Scientific Instruments},
  volume  = {80},
  number  = {5},
  pages   = {053702},
  year    = {2009},
  doi     = {10.1063/1.3127589}
}

@article{Heinrich2004SpinFlip,
  author  = {Heinrich, A. J. and Gupta, J. A. and Lutz, C. P. and Eigler, D. M.},
  title   = {Single-Atom Spin-Flip Spectroscopy},
  journal = {Science},
  volume  = {306},
  number  = {5695},
  pages   = {466--469},
  year    = {2004},
  doi     = {10.1126/science.1101077}
}

@article{Hirjibehedin2006SpinCoupling,
  author  = {Hirjibehedin, Cyrus F. and Lutz, Christopher P. and Heinrich, Andreas J.},
  title   = {Spin Coupling in Engineered Atomic Structures},
  journal = {Science},
  volume  = {312},
  number  = {5776},
  pages   = {1021--1024},
  year    = {2006},
  doi     = {10.1126/science.1125398}
}

@article{Seifert2020,
	author = {Seifert, T. S. and Kovarik, S. and Nistor, C. and Persichetti, L. and Stepanow, S. and Gambardella, P.},
	year = {2020},
	title = {Single-atom electron paramagnetic resonance in a scanning tunneling microscope driven by a radio-frequency antenna at 4\,{K}},
	volume = {2},
	number = {1},
	pages = {013032},
	journal = {Physical Review Research},
	doi = {10.1103/PhysRevResearch.2.013032}
}

@article{MartinezBlanco2015Gating,
  author  = {Mart\'inez-Blanco, Jes\'us and Nacci, Christophe and Erwin, Steven C. and Kanisawa, Kiyoshi and Locane, Elina and Thomas, Mark and von Oppen, Felix and Brouwer, Piet W. and F\"olsch, Stefan},
  title   = {Gating a single-molecule transistor with individual atoms},
  journal = {Nature Physics},
  volume  = {11},
  pages   = {640--644},
  year    = {2015},
  doi     = {10.1038/nphys3385}
}

@article{Misra2013STM,
  author  = {Misra, S. and Zhou, B. B. and Drozdov, I. K. and Seo, J. and Urban, L. and Gyenis, A. and Kingsley, S. C. J. and Jones, H. and Yazdani, A.},
  title   = {Design and performance of an ultra-high vacuum scanning tunneling microscope operating at dilution refrigerator temperatures and high magnetic fields},
  journal = {Review of Scientific Instruments},
  volume  = {84},
  number  = {10},
  pages   = {103903},
  year    = {2013},
  doi     = {10.1063/1.4822271}
}

@article{Esat2021MillikelvinSTM,
  author  = {Esat, Taner and Borgens, Peter and Yang, Xiaosheng and Coenen, Peter and Cherepanov, Vasily and Raccanelli, Andrea and Tautz, F. Stefan and Temirov, Ruslan},
  title   = {A millikelvin scanning tunneling microscope in ultra-high vacuum with adiabatic demagnetization refrigeration},
  journal = {Review of Scientific Instruments},
  volume  = {92},
  number  = {6},
  pages   = {063701},
  year    = {2021},
  doi     = {10.1063/5.0050532}
}

@article{Song2010,
  author  = {Song, Young Jae and Otte, Alexander F. and Shvarts, Vladimir and Zhao, Zuyu and Kuk, Young and Blankenship, Steven R. and Band, Alan and Hess, Frank M. and Stroscio, Joseph A.},
  journal = {Review of Scientific Instruments},
  title   = {Invited Review Article: A 10 {mK} scanning probe microscopy facility},
  year    = {2010},
  number  = {12},
  pages   = {121101},
  volume  = {81},
  doi     = {10.1063/1.3520482}
}

@article{vonAllworden2018,
  author  = {von Allw\"orden, Henning and Eich, Andreas and Knol, Elze J. and Hermenau, Jan and Sonntag, Andreas and Gerritsen, Jan W. and Wegner, Daniel and Khajetoorians, Alexander A.},
  journal = {Review of Scientific Instruments},
  title   = {Design and performance of an ultra-high vacuum spin-polarized scanning tunneling microscope operating at 30 {mK} and in a vector magnetic field},
  year    = {2018},
  number  = {3},
  pages   = {033902},
  volume  = {89},
  doi     = {10.1063/1.5020045}
}

@article{Machida2018,
  author  = {Machida, T. and Kohsaka, Y. and Hanaguri, T.},
  journal = {Review of Scientific Instruments},
  title   = {A scanning tunneling microscope for spectroscopic imaging below 90 {mK} in magnetic fields up to 17.5 {T}},
  year    = {2018},
  number  = {9},
  pages   = {093707},
  volume  = {89},
  doi     = {10.1063/1.5049619}
}

@article{Wong2020,
  author  = {Wong, Dillon and Jeon, Sangjun and Nuckolls, Kevin P. and Oh, Myungchul and Kingsley, Simon C. J. and Yazdani, Ali},
  journal = {Review of Scientific Instruments},
  title   = {A modular ultra-high vacuum millikelvin scanning tunneling microscope},
  year    = {2020},
  number  = {2},
  pages   = {023703},
  volume  = {91},
  doi     = {10.1063/1.5132872}
}

@article{Assig2013,
  author  = {Assig, Maximilian and Etzkorn, Markus and Enders, Axel and Stiepany, Wolfgang and Ast, Christian R. and Kern, Klaus},
  journal = {Review of Scientific Instruments},
  title   = {A 10 {mK} scanning tunneling microscope operating in ultra high vacuum and high magnetic fields},
  year    = {2013},
  number  = {3},
  pages   = {033903},
  volume  = {84},
  doi     = {10.1063/1.4793793}
}

@article{Roychowdhury2014,
  author  = {Roychowdhury, Anita and Gubrud, M. A. and Dana, R. and Anderson, J. R. and Lobb, C. J. and Wellstood, F. C. and Dreyer, M.},
  journal = {Review of Scientific Instruments},
  title   = {A 30 {mK}, 13.5 {T} scanning tunneling microscope with two independent tips},
  year    = {2014},
  number  = {4},
  pages   = {043706},
  volume  = {85},
  doi     = {10.1063/1.4871056}
}

@article{Schwenk2020,
  author  = {Schwenk, Johannes and Kim, Sungmin and Berwanger, Julian and Ghahari, Fereshte and Walkup, Daniel and Slot, Marlou R. and Le, Son T. and Cullen, William G. and Blankenship, Steven R. and Vranjkovic, Sasa and Hug, Hans J. and Kuk, Young and Giessibl, Franz J. and Stroscio, Joseph A.},
  journal = {Review of Scientific Instruments},
  title   = {Achieving $\mu${eV} tunneling resolution in an \textit{in-operando} scanning tunneling microscopy, atomic force microscopy, and magnetotransport system for quantum materials research},
  year    = {2020},
  number  = {7},
  pages   = {071101},
  volume  = {91},
  doi     = {10.1063/5.0005320}
}

@article{hurdax_integer_2025,
  author  = {Hurdax, Philipp and Hollerer, Michael and Kern, Christian S. and Puschnig, Peter and Sterrer, Martin and Ramsey, Michael G.},
  title   = {Integer Charge Transfer Model---{PTCDA} on {MgO}(001)/{Ag}(001) Probing the Transition from Single to Double Integer Charge Transfer},
  journal = {The Journal of Physical Chemistry C},
  volume  = {129},
  number  = {2},
  pages   = {1553--1561},
  year    = {2025},
  doi     = {10.1021/acs.jpcc.4c08104}
}

@article{hurdax_large_2022,
  author  = {Hurdax, Philipp and Kern, Christian S. and Bon{\'e}, Thomas Georg and Haags, Anja and Hollerer, Michael and Egger, Larissa and Yang, Xiaosheng and Kirschner, Hans and Gottwald, Alexander and Richter, Mathias and Bocquet, Fran{\c{c}}ois C. and Soubatch, Serguei and Koller, Georg and Tautz, Frank Stefan and Sterrer, Martin and Puschnig, Peter and Ramsey, Michael G.},
  title   = {Large Distortion of Fused Aromatics on Dielectric Interlayers Quantified by Photoemission Orbital Tomography},
  journal = {ACS Nano},
  volume  = {16},
  number  = {10},
  pages   = {17435--17443},
  year    = {2022},
  doi     = {10.1021/acsnano.2c08631}
}

@article{HollandSteckelmacher1952,
  author  = {Holland, L. and Steckelmacher, W.},
  title   = {The Distribution of Thin Films Condensed on Surfaces by the Vacuum Evaporation Method},
  journal = {Vacuum},
  volume  = {2},
  number  = {4},
  pages   = {346--364},
  year    = {1952},
  doi     = {10.1016/0042-207X(52)93784-6}
}

@book{Ohring2002,
  author    = {Ohring, Milton},
  title     = {Materials Science of Thin Films: Deposition and Structure},
  edition   = {2},
  publisher = {Academic Press},
  address   = {San Diego},
  year      = {2002},
  isbn      = {978-0-12-524975-1}
}

@article{Baumann2015,
  author    = {Baumann, S. and Donati, F. and Stepanow, S. and Rusponi, S. and Paul, W. and Gangopadhyay, S. and Rau, I. G. and Pacchioni, G. E. and Gragnaniello, L. and Pivetta, M. and Dreiser, J. and Piamonteze, C. and Lutz, C. P. and Macfarlane, R. M. and Jones, B. A. and Gambardella, P. and Heinrich, A. J. and Brune, H.},
  title     = {Origin of Perpendicular Magnetic Anisotropy and Large Orbital Moment in {Fe} Atoms on {MgO}},
  journal   = {Physical Review Letters},
  volume    = {115},
  number    = {23},
  pages     = {237202},
  year      = {2015},
  publisher = {American Physical Society},
  doi       = {10.1103/PhysRevLett.115.237202}
}

\end{document}